\documentclass[preprint,12pt]{elsarticle}

\usepackage{amssymb}
\usepackage{amsmath}
\usepackage{amsfonts}
\usepackage{amssymb}
\usepackage{graphicx}
\usepackage{multirow}
\usepackage{setspace}
\usepackage{amssymb}
\usepackage{hyperref}
\usepackage{booktabs}
\usepackage{multirow}
\usepackage{graphicx}
\usepackage{float}
\usepackage{tabulary,graphicx,times,fancyhdr,amsfonts,amssymb,amsbsy,latexsym,amsmath,array}
\usepackage{url,multirow,morefloats,floatflt,cancel,tfrupee,textcomp,colortbl,xcolor,pifont}
\usepackage{tabularx}
\usepackage[nointegrals]{wasysym}
\usepackage{cleveref}
\usepackage{amsthm}
\usepackage{mathptmx}
\usepackage{mathrsfs}
\usepackage[graphicx]{realboxes}
\usepackage{hyperref}
\usepackage{algorithm}
\usepackage{algorithmic}
\usepackage{amsmath}
\usepackage{amssymb}
\usepackage{tocbibind}
\usepackage{lipsum} 
\usepackage{forloop} 
\usepackage[switch]{lineno}

\usepackage{makecell}
\usepackage{cleveref}
\usepackage[utf8]{inputenc}
\usepackage[footnotesize,labelfont={sc,bf,normalsize},font={footnotesize},format=plain]{caption}
\usepackage[left=1in,top=1in,right=1in,bottom=1in]{geometry}
\usepackage{subcaption}

\begin{document}

\begin{frontmatter}



\title{QTMRL: An Agent for Quantitative Trading Decision-Making Based on Multi-Indicator Guided Reinforcement Learning} 

\author[label1]{Jinfeng Pan} 
\author[label2]{Jiahao Chen} 

\affiliation[label1]{
    organization={Guangdong University of Finance},  
    city={Guangzhou},  
    state={Guangdong},  
    country={China}  
}

\affiliation[label2]{
	organization={Jinan University},  
	city={Guangzhou},  
	state={Guangdong},  
	country={China}  
}

\begin{abstract}
In the highly volatile and uncertain global financial markets, traditional quantitative trading models relying on statistical modeling or empirical rules often fail to adapt to dynamic market changes and black swan events due to rigid assumptions and limited generalization. To address these issues, this paper proposes QTMRL (Quantitative Trading Multi-Indicator Reinforcement Learning), an intelligent trading agent that combines multi-dimensional technical indicators with reinforcement learning (RL) for adaptive and stable portfolio management. We first construct a comprehensive multi-indicator dataset using 23 years of S\&P 500 daily OHLCV data (2000–2022) for 16 representative stocks across 5 sectors, enriching raw data with trend, volatility, and momentum indicators to capture holistic market dynamics. Then, we design a lightweight RL framework based on the Advantage Actor-Critic (A2C) algorithm, which includes data processing, A2C algorithm, and trading agent modules to support policy learning and actionable trading decisions. Extensive experiments compare QTMRL with 9 baselines (e.g., ARIMA, LSTM, moving average strategies) across diverse market regimes, verifying its superiority in profitability, risk adjustment, and downside risk control. The code of QTMRL is publicly available at \url{https://github.com/ChenJiahaoJNU/QTMRL.git.}
\end{abstract}

%


\end{frontmatter}


\newpage
\section{Introduction}
In today’s highly volatile and uncertain global financial markets, managing individual stock portfolios for stable returns remains a key concern for academia and practitioners. Traditional methods, based on statistical modeling or empirical rules, often fail amid dynamic market shifts and black swan events, as their assumptions grow disconnected from reality, blocking universally applicable frameworks. With AI’s deep integration in finance, these technologies offer novel pathways to overcome conventional limitations. This sparks a critical question: Can we build an "investment guru-like" intelligent agent to autonomously manage portfolios for stable returns?


Historically, a multitude of models for quantitative trading and portfolio management have emerged over time. Prior to the rise of deep learning, approaches such as moving averages, trend indicators, and predictive models based on artificial neural networks (ANN) were widely applied in quantitative trading practices \cite{before20121,before20122,before20123}. With the widespread adoption of deep learning and Python in financial domains, more advanced models including long short-term memory (LSTM), recurrent neural networks (RNN), convolutional neural networks (CNN), and autoregressive integrated moving average (ARIMA) gained prominence in quantitative trading research and applications \cite{before20221,before20223}. Following the explosive popularity of large language models (LLMs) like ChatGPT, there has been a growing body of research exploring the application of large-scale language models in quantitative trading and portfolio management \cite{LLM4fin1,LLM4fin2,LLM4fin3}. 

%
%
%
%
%
The extensive body of related research underscores the inherent challenges in developing quantitative trading models, with three key hurdles standing out: First, stock investment requires real-time monitoring of multiple stock indicators and making buy-sell decisions across a diverse set of target stocks, demanding simultaneous processing of multi-dimensional information and dynamic decision-making. Second, a sophisticated investor or agent must possess strong generalization capabilities, enabling effective management of arbitrary stocks based on accumulated investment experience. Third, a robust investor or agent should be able to mitigate losses or even generate returns amid black swan events, which are characterized by extreme market volatility and unpredictability \cite{before20222}.

To address these challenges, we propose an Agent for Quantitative Trading Decision-Making Based on Multi-Indicator Guided Reinforcement Learning (QTMRL). The following key features characterize this novel framework: First, unlike traditional models, we leverage a diverse set of indicators to provide the agent with comprehensive evidence for investment decision-making, ensuring a multi-faceted understanding of market dynamics. Second, we employ a reinforcement learning (RL)-based agent, which possesses the ability to continuously learn and reflect from failures, thereby fostering robust generalization capabilities across different market scenarios. Third, the reinforcement learning algorithm we adopt offers a key advantage: its optimization framework effectively mitigates overfitting, enabling the agent to maintain performance advantages even during black swan events. To this end, the primary contributions of our paper are as follows:

\begin{itemize}
	\item Collection and integration of diverse technical indicators for stock market analysis, constructing a multi-dimensional dataset to support the agent’s comprehensive market assessment.  
	
	\item Development of a novel, lightweight reinforcement learning framework for quantitative trading, specifically tailored to empower the agent with efficient decision-making capabilities in dynamic market environments.  
	
	\item Experimental validation across bull and bear markets, demonstrating the agent’s risk mitigation abilities through comparative performance analysis under distinct market conditions.  
\end{itemize}

\section{Related Work}

\subsection{Quantitative Trading}
Since the 1970s, quantitative trading (QT) has emerged as a prominent focus in both academic research and the financial industry. In the United States—home to the world’s most developed financial sector—quantitative trading now accounts for over 70\% of total stock market turnover \cite{QT1}. With the rapid advancement of modern statistical and mathematical methodologies, quantitative trading leverages computer technology to execute trading and investment activities. Through this approach, investors can assess the probabilities of various market trends and automate trading decisions via programmed instructions to implement investment transactions \cite{QT2}.  

The quantitative trading process can be conceptualized as an online decision-making strategy \cite{QT3, QT4}, encompassing market condition analysis and the execution of optimal actions \cite{QT5}. However, accurately predicting stock price trends and executing trades remains extremely challenging when using traditional machine learning algorithms, primarily because financial market dynamics are influenced by a complex interplay of numerous external factors. However, our research takes a reinforcement learning perspective, focusing on constructing an intelligent agent capable of autonomously learning quantitative trading strategies. By fully leveraging the reflective characteristics inherent in such models, our goal is to develop a quantitative trading framework with sufficient generalizability across diverse market scenarios.


\subsection{Reinforcement Learning}


Reinforcement Learning (RL), a pivotal branch of machine learning, focuses on how intelligent agents learn optimal behavioral strategies through interactions with their environment to maximize cumulative rewards \cite{RL1}.  Classical RL has gradually developed a rigorous mathematical foundation based on Markov Decision Processes (MDPs), giving rise to fundamental algorithms such as dynamic programming, Monte Carlo methods, and temporal difference (TD) learning \cite{RL2, REINFORCE}. With the rise of deep learning, deep reinforcement learning has enabled end-to-end training of perception and decision-making systems. A notable breakthrough was DeepMind's Deep Q-Network (DQN), which mitigates the instability of value function approximation through the use of experience replay and target networks \cite{DQN}.  Concurrently, policy gradient methods such as REINFORCE and Actor-Critic frameworks like A2C and A3C have been introduced, enabling direct optimization of policy functions \cite{A2C, A3C}. These advancements have driven RL to achieve significant success in complex tasks such as Atari gameplay and the board game Go.

Take it simple, reinforcement learning (RL) is about training an intelligent agent to optimize its own strategies automatically. Its key advantage lies in the fact that the model can select optimal actions based solely on reward values, without relying on the accuracy metrics required in traditional supervised learning. Currently, reinforcement learning has found numerous applications in quantitative trading. However, research on open-source, multi-factor-driven reinforcement learning models for quantitative trading remains relatively scarce—and this is precisely the primary goal of our research. 
\section{Preliminary}

Reinforcement Learning (RL) enables an intelligent agent to acquire optimal strategies through interaction with a dynamic environment. This framework is particularly well-suited for quantitative trading, where agents must adapt to market fluctuations in real time. RL is generally characterized by five essential components. The \textit{environment} corresponds to the financial market, which provides multi-asset price dynamics and trading rules. The \textit{state} represents market conditions at time $t$, constructed from a time window of technical indicators for each asset together with the current portfolio status. Formally, the state is denoted as $s_t \in \mathbb{R}^{W \times N \times F}$, where $W$ is the window size, $N$ the number of assets, and $F$ the number of features. The \textit{action} space contains discrete trading decisions across multiple assets. Each action is encoded as an integer representing a combination of buy and sell operations, resulting in $2^N$ possible actions. The \textit{reward} is a scalar feedback signal that reflects trading performance, including profits, portfolio growth, and penalties for invalid actions. Finally, the \textit{policy} is a neural network $\pi_\theta(a|s)$ that maps states to probabilities of actions, optimized to maximize the cumulative discounted return
$$
G_t = \sum_{k=0}^\infty \gamma^k r_{t+k},
$$
where $\gamma$ is the discount factor.

Quantitative trading can be formulated as a Markov Decision Process (MDP) defined by the tuple $(\mathcal{S}, \mathcal{A}, P, R, \gamma)$. Here, $\mathcal{S}$ denotes the state space, $\mathcal{A}$ the action space, $P(s'|s,a)$ the transition dynamics, $R(s,a)$ the reward function, and $\gamma$ the discount factor. The Markov property ensures that the next state depends only on the current state and action, which aligns well with the temporal nature of financial markets.

Our framework employs the Advantage Actor-Critic (A2C) algorithm. This method integrates two neural networks: the \textit{actor} $\pi_\theta$, which outputs probabilities for exploring different trading actions, and the \textit{critic} $V_\phi$, which estimates state values to assess the quality of these actions. The optimization relies on the advantage function
$$
A(s,a) = Q(s,a) - V(s),
$$
which measures how much better an action is compared to the expected value of the state. The overall training objective combines policy, value, and entropy losses:
$$
\mathcal{L}_{\text{total}} = \mathcal{L}_{\text{policy}} + c_v \cdot \mathcal{L}_{\text{value}} + c_e \cdot \mathcal{L}_{\text{entropy}}.
$$

Training proceeds by collecting trajectories from interactions with the environment, computing returns and advantages, and iteratively updating the networks. This process gradually refines the learned policy to improve trading performance, as outlined in Algorithm~\ref{alg:a2c_trading}.

\section{Methodology}
\subsection{Datasets Construction}

We constructed our dataset from the S\&P 500 daily timeseries dataset, focusing on 16 representative stocks across multiple sectors. Raw OHLCV data was preprocessed by standardizing columns, sorting chronologically, forward-filling missing prices, and replacing missing volumes with zeros.

To enrich raw market data, we computed multiple categories of technical indicators:

\begin{itemize}
\item Trend indicators: simple and exponential moving averages (SMA, EMA), Heiken Ashi (HA), and Ichimoku Cloud, capturing long–short horizon trends.
\item Volatility indicators: standard deviation (STDDEV), Average True Range (ATR), Bollinger Bands (BBands), quantifying market uncertainty.
\item Momentum indicators: Relative Strength Index (RSI), Moving Average Convergence Divergence (MACD), and SuperTrend, reflecting overbought/oversold conditions and momentum shifts.

\end{itemize}

Each stock $i$ at time $t$ is represented as a feature vector

$$
x_t^i = \big[ \text{OHLCV}_t^i, \ \text{Trend}_t^i, \ \text{Volatility}_t^i, \ \text{Momentum}_t^i \big] \in \mathbb{R}^F ,
$$

where $F$ is the number of features. Aggregating over $N$ assets and a rolling window of length $W$, the reinforcement learning state is

$$
s_t = \{ x_{t-W+1}^1, \dots, x_t^1; \ \dots \ ; x_{t-W+1}^N, \dots, x_t^N \} \in \mathbb{R}^{W \times N \times F}.
$$

This construction yields a multi-dimensional dataset that integrates raw OHLCV with diverse technical indicators, providing the RL agent with both low-level price dynamics and high-level market signals.

\subsection{Framework}

\begin{figure}[h]
	\centering
	\includegraphics[width=\linewidth]{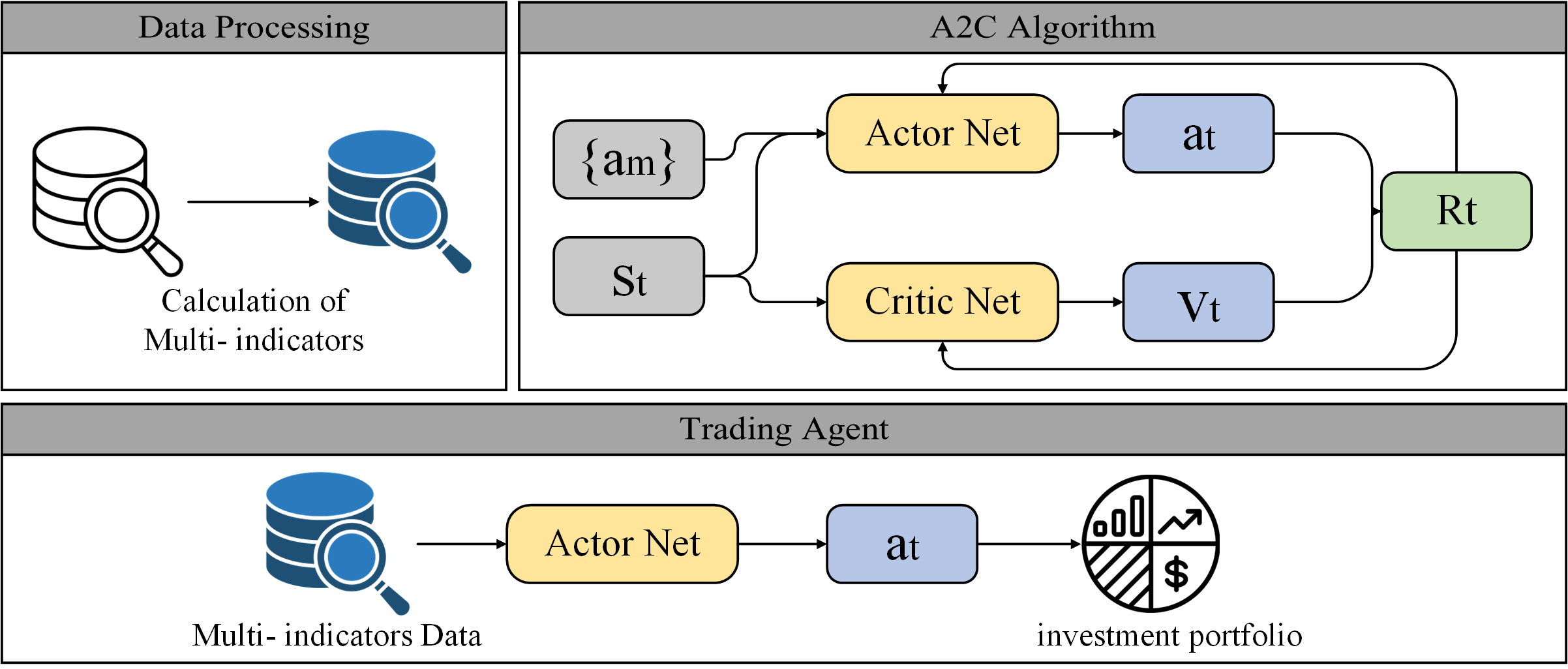}
	\caption{Framework for QTMRL}
	\label{fig:framework}
\end{figure}

The QTMRL framework integrates three interconnected modules: data processing, reinforcement learning via the A2C algorithm, and trading execution. Each module is responsible for a distinct yet complementary function, as illustrated in Fig.~\ref{fig:framework}, ensuring a systematic transformation from raw market information to actionable trading strategies.
\subsubsection{Data Processing Module }
The data processing module serves as the foundation of the framework by converting raw financial market data into enriched multi-indicator datasets. The raw inputs, typically represented as OHLCV (Open, High, Low, Close, Volume), are transformed into a comprehensive set of technical indicators that capture various aspects of market behavior. These include trend indicators (e.g., SMA, EMA), volatility measures (e.g., ATR, Bollinger Bands), momentum oscillators (e.g., RSI, stochastic indicators), and other advanced features such as Ichimoku components (Tenkan Sen, Kijun Sen, Senkou Spans) and Heikin-Ashi transformations. Examples of features generated in this step include ``ATR\_10'', ``RSI\_14'', ``EMA\_26'', ``SMA\_50'', and ``macd\_12\_26\_9''. By standardizing and expanding the raw data into a multi-dimensional feature space, this module enables the reinforcement learning agent to interpret market states more comprehensively and facilitates the learning of patterns that are not directly observable from raw price and volume data.

\subsubsection{A2C Algorithm Module }
The reinforcement learning core of QTMRL is implemented using the Advantage Actor-Critic (A2C) algorithm, which balances exploration and evaluation in decision-making. In this module, the processed multi-indicator data forms the state representation \(S_t\), encapsulating the prevailing market condition at time \(t\). The Actor network parameterizes the policy \(\pi_\theta(a|s)\), producing a probability distribution over the set of admissible trading actions \(\{a_m\}\), such as buying, selling, or holding assets. The Critic network, on the other hand, estimates the value function \(V_t\), representing the expected cumulative reward from state \(S_t\). Together, these two components allow the computation of the advantage function \(A(s,a)=Q(s,a)-V(s)\), which quantifies the relative benefit of taking a particular action in a given state. The reward signal \(R_t\), derived from trading performance metrics such as profit and portfolio growth, guides the joint optimization of policy and value networks by minimizing policy loss, value loss, and entropy loss. Through this iterative process, the A2C module equips the trading agent with the capacity to learn robust strategies that adapt to dynamic market conditions.

\subsubsection{Trading Agent Module }
The trading agent module operationalizes the learned policy into actionable investment decisions. It takes as input the processed multi-indicator data, identical to the state representation \(S_t\) in the A2C module, and employs the pre-trained Actor network to determine the optimal action \(a_t\) under current market conditions. This decision, which may involve buying, selling, or rebalancing asset positions, is subsequently executed within the investment portfolio, as represented by the portfolio icon in Fig.~\ref{fig:framework}. The module thereby serves as the interface between reinforcement learning outputs and real-world portfolio management, ensuring that theoretical policy optimization is translated into practical trading performance. Its ultimate objective is to enable adaptive responses across market regimes—whether bullish, bearish, or volatile—and to maintain stable, risk-adjusted returns over time.

\section{Experience}
\subsection{Experimental}
\subsubsection{Datasets}

The experimental data is derived from the publicly accessible \textit{jwigginton/timeseries-daily-sp500} dataset hosted on Hugging Face, which provides daily records of Open, High, Low, Close, and Volume (OHLCV) metrics for constituent stocks of the S\&P 500 index. To ensure comprehensive coverage of diverse market dynamics and industry-specific volatility profiles, we selected 16 representative stocks across five key economic sectors: cruise lines (Carnival Corporation [CCL], Royal Caribbean Group [RCL]); airlines (American Airlines Group [AAL], United Airlines Holdings [UAL], Delta Air Lines [DAL]); energy (Marathon Oil Corporation [MRO], Occidental Petroleum Corporation [OXY]); casino and resort (Wynn Resorts [WYNN], Las Vegas Sands Corporation [LVS]); and technology/finance (Apple Inc. [AAPL], American Express Co [AXP], Bank of America Corp [BAC], Chevron Corp [CVX], Alphabet Inc. [GOOGL], Johnson \& Johnson [JNJ], Microsoft Corp [MSFT], NVIDIA Corp [NVDA]).  

The subset of the dataset used in our experiments spans a 23-year period from January 3, 2000, to December 30, 2022, encompassing multiple distinct market regimes—including prolonged bullish expansions, acute bearish downturns, and transitional volatile phases—and critical historical events such as the 2008 global financial crisis, the 2020 COVID-19 pandemic-induced market crash, and subsequent recovery periods. This extensive temporal scope enables rigorous evaluation of the model’s adaptability across both normal market conditions and extreme systemic shocks. Raw data underwent systematic preprocessing: missing price values were forward-filled to preserve temporal continuity, zero-volume entries were retained to reflect illiquid market periods, and all features were standardized using z-score normalization to eliminate scale biases, ensuring compatibility with the reinforcement learning framework’s input requirements.

\subsection{Evaluation Metrics}
To comprehensively assess the proposed trading strategy, we adopt four standard quantitative metrics to measure profitability, risk-adjusted returns, and downside risk\cite{EvaluationMetrics}. All metrics are calculated based on the average price of the investment portfolio, and each metric value is derived from the average of the corresponding metrics across multiple stocks in the portfolio:

\begin{itemize}
	\item \textbf{Total Return Rate (Tr)}: Measures overall profitability as the relative change in total portfolio value (positions + cash). It is calculated as the average of total return rates across all selected stocks in the portfolio:
	$$
	Tr = \frac{P_{\text{end}} - P_{\text{start}}}{P_{\text{start}}}
	$$
	where the final portfolio value \(P_{\text{end}}\) and initial portfolio value \(P_{\text{start}}\) are derived from the average performance of multiple stocks.
	
	\item \textbf{Sharpe Ratio (Sr)}: Quantifies risk-adjusted returns (Sharpe, 1966) as the ratio of expected returns to their volatility. It is computed as the average of Sharpe ratios calculated for each individual stock in the portfolio:
	$$
	Sr = \frac{E[r]}{\sigma[r]}
	$$
	where \(E[r]\) is the average expected return across stocks, and \(\sigma[r]\) is the average volatility of returns across stocks.
	
	\item \textbf{Volatility (Vol)}: Represented by \(\sigma[r]\), the standard deviation of historical returns, reflecting return variability and associated risks. It is determined as the average volatility of returns across all stocks in the portfolio.
	
	\item \textbf{Maximum Drawdown (Mdd)}: Measures the worst-case loss as the maximum peak-to-trough decline. It is calculated as the average of maximum drawdown values across all stocks in the portfolio:
	$$
	Mdd = \max\left( \frac{P_i - P_j}{P_i} \right) \quad \text{for } j > i
	$$
	where \(P_i\) and \(P_j\) are peak and trough portfolio values derived from the average price performance of multiple stocks.
\end{itemize}

Specifically, in our implementation, after calculating each metric for individual stocks (including daily returns, volatility, and drawdown), we aggregate these metrics by taking the mean across all selected stocks in the portfolio for each evaluation period (2019–2021). This approach ensures that the reported metrics reflect the overall performance of the multi-stock portfolio rather than individual stock behavior.

\subsubsection{Training Details}
The training process adopted a consistent experimental setup across all compared models (including our QTMRL and seven baselines: ARIMA, LSTM, CNN, ANN, Random Strategy, 10/20/30-day Moving Average (MA) strategies, and Dow Jones Tracking Strategy) to ensure fair comparison, with all models leveraging the same 23-year S\&P 500 subset (January 3, 2000–December 30, 2022) as the foundational dataset—consistent with the preprocessed data (forward-filled missing values, retained zero-volume entries, z-score standardized features) described earlier. Models were trained on ten years of historical data from this subset and evaluated on the 2020 test period (January 3–December 30), while the broader experimental framework covered 2019–2021 for robustness; all shared a unified multi-asset trading environment with core hyperparameters: \$10,000 initial capital, 0.05\% transaction fee rate, 20-step window size for feature construction, 20\% of available capital per buy (shares purchased = (current capital × 0.2)/current stock price), 50\% of current holdings per sell (shares sold = current position × 0.5), and a global random seed of 42 (with additional seeds 43–46 for Random Strategy stability checks).

Baseline-specific configurations and training environments were standardized as follows: The Random Strategy used fixed action probabilities (20\% buy, 20\% sell, 60\% hold) across 5 seed trials; MA strategies generated signals via price-MA crossovers (T=10/20/30); the Dow Jones Tracking Strategy relied on quarterly/year-end rebalancing; ARIMA adopted (5,1,0) order parameters with a 0.5\% price change threshold; LSTM/CNN/ANN used Adam optimizer (0.0001 learning rate), cross-entropy loss, 100-step episodes, and 1M training timesteps; A2C used Adam optimizer (0.00001 learning rate), 50-step rollouts, $\gamma$=0.96, entropy coefficient=0.05, and 1M training timesteps. All training was conducted on an 8×NVIDIA A800 GPU cluster, with deep learning models converging in 2 hours. Post-training performance was evaluated using the four portfolio-level metrics (Total Return Rate, Sharpe Ratio, Volatility, Maximum Drawdown) detailed earlier—aggregated as the mean of individual stock metrics across the 16 selected stocks to reflect multi-asset portfolio performance.

\subsection{Results}
\begin{table}[H]
	\centering
	\caption{Overall Performance Comparison Between QTMRL and Baseline Models (Test Set: 2020)}
	\label{tab:2020Performance}
	\resizebox{\textwidth}{!}{%
		\begin{tabular}{cccccc}
			\rowcolor[gray]{0.8} 
	\hline
	Year & Strategy & Return\_Rate & Sharpe\_Ratio & Volatility & Max\_Drawdown \\
	\hline
	2020 & Random Strategy & -0.03 & 0.38 & 53.01 & -39.06 \\
	\hline
	2020 & Dow Jones Tracking Strategy & -2.55 & 0.34 & 72.58 & -55.12 \\
	\hline
	2020 & 10-Day Moving Average Strategy & 8.38 & 0.42 & 45.75 & -34.23 \\
	\hline
	2020 & 20-Day Moving Average Strategy & 11.88 & 0.4 & 40.46 & -35.52 \\
	\hline
	2020 & 30-Day Moving Average Strategy & 8.54 & 0.37 & 39.01 & -35.05 \\
	\hline
2020 & ARIMA & & & &\\
\hline
2020 & LSTM & & & &\\
\hline
2020 & CNN & & & &\\
\hline
2020 & ANN & & & &\\
\hline
2020 & A2C & & & &\\
\hline
2020 & QTMRL & & & &\\
\end{tabular}
}
\end{table}

\begin{table}[H]
	\centering
	\caption{Overall Performance Comparison Between QTMRL and Baseline Models (Test Set: 2021)}
	\label{tab:2021Performance}
	\resizebox{\textwidth}{!}{%
		\begin{tabular}{cccccc}
			\rowcolor[gray]{0.8} 
			\hline
	Year & Strategy & Return\_Rate & Sharpe\_Ratio & Volatility & Max\_Drawdown \\
	\hline
	2021 & Random Strategy & 23.18 & 0.91 & 24.8 & -21.05 \\
	\hline
	2021 & Dow Jones Tracking Strategy & 46 & 1.23 & 35.42 & -23.11 \\
	\hline
	2021 & 10-Day Moving Average Strategy & 28.11 & 1.15 & 24.51 & -16.58 \\
	\hline
	2021 & 20-Day Moving Average Strategy & 26.6 & 1.08 & 24.59 & -16.67 \\
	\hline
2021 & 30-Day Moving Average Strategy & 11.39 & 0.52 & 20.15 & -16 \\
\hline
2021 & ARIMA & & & &\\
\hline
2021 & LSTM & & & &\\
\hline
2021 & CNN & & & &\\
\hline
2021 & ANN & & & &\\
\hline
2021 & A2C & & & &\\
\hline
2021 & QTMRL & & & &\\
\end{tabular}
}
\end{table}
\subsubsection{Overall Performance}

\subsubsection{Ablation Study}

\subsubsection{Visualization}

\section{Conclution}

\newpage

\bibliographystyle{elsarticle-num-names} 
\bibliography{reference}

\appendix

\section{Algorithm}
\begin{algorithm}[H]
	\caption{A2C Algorithm for Multi-Asset Trading}
	\label{alg:a2c_trading}
	\begin{algorithmic}[1] 
		\STATE Initialize actor $\pi_\theta$ and critic $V_\phi$ networks; 
		\STATE set optimizers and hyperparameters $\gamma, c_v, c_e$;
		
		\FOR{each training episode}
		\STATE Reset environment to get initial state $\mathbf{s}_0$;
		\STATE Collect trajectory $\{(\mathbf{s}_t, \mathbf{a}_t, \mathbf{r}_t, \mathbf{s}_{t+1})\}$ over $T$ timesteps:
		\STATE \quad $\mathbf{a}_t \sim \pi_\theta(\cdot|\mathbf{s}_t)$;
		\STATE \quad $\mathbf{s}_{t+1}, \mathbf{r}_t, \text{done} = \text{environment.step}(\mathbf{a}_t)$;
		\STATE \quad store $(\mathbf{s}_t, \mathbf{a}_t, \mathbf{r}_t)$;
		
		\STATE Compute returns and advantages:
		\STATE \quad $\mathbf{G}_T = 0$ if $\text{done}$ else $V_\phi(\mathbf{s}_T)$;
		\STATE \quad \FOR{$t = T-1$ downto $0$}
		\STATE \quad \quad $\mathbf{G}_t = \mathbf{r}_t + \gamma \cdot \mathbf{G}_{t+1}$;
		\STATE \quad \quad $V_t = V_\phi(\mathbf{s}_t)$;
		\STATE \quad \quad $\mathbf{A}_t = \mathbf{G}_t - V_t$;
		\STATE \quad \ENDFOR
		
		\STATE Update networks:
		\STATE \quad $\text{policy\_loss} = -\mathbb{E}[\log \pi_\theta(\mathbf{a}_t|\mathbf{s}_t) \cdot \mathbf{A}_t]$;
		\STATE \quad $\text{value\_loss} = \mathbb{E}[(V_\phi(\mathbf{s}_t) - \mathbf{G}_t)^2]$;
		\STATE \quad $\text{entropy\_loss} = -\mathbb{E}[H(\pi_\theta(\cdot|\mathbf{s}_t))]$;
		\STATE \quad $\text{total\_loss} = \text{policy\_loss} + c_v \cdot \text{value\_loss} + c_e \cdot \text{entropy\_loss}$;
		\STATE \quad Update $\theta$ and $\phi$ by minimizing total\_loss;
		\ENDFOR
		
		\STATE Return trained policy $\pi_\theta$;
	\end{algorithmic}
\end{algorithm}
\end{document}